\documentclass[a4paper, times, 11pt,latex8]{article} 

\usepackage{amsmath}
\usepackage{amssymb}
\usepackage{amsfonts}
\usepackage{amstext}
\usepackage{latexsym}
\usepackage{graphicx}

{}{}
\newtheorem{thm}{Theorem}{}{}
\newtheorem{lem}{Lemma}{}{}
{}{}
{}{}
{}{} 
{}{}

\newcommand{\commentout}[1]{}

\newcommand{\intp}[1]{{[#1]}_{\pi}}

\newcommand{\ket}[1]{|#1\rangle}

\newcommand{\bra}[1]{\langle #1|}

\newcommand{\inpr}[3]{\langle #1|#2|#3\rangle} 
\newcommand{\inpop}[2]{\langle #1,#2\rangle}

\newcommand{\hconj}[1]{{#1}^{\dagger}} 
 
\newcommand{\tr}{{\tt Tr}} 
\newcommand {\pj}[1]{\ket{#1}\bra{#1}}

\newcommand{\bcolvec}{\left(\begin{array}{c}}
\newcommand{\ecolvec}{\end{array}\right)}
\newcommand{\bmat}[1]{\left(\begin{array}{#1}}
\newcommand{\emat}{\end{array}\right)}

\newcommand{\imp}{\Rightarrow}

\newcommand{\pdist}[1]{p(#1)}
\newcommand{\rvec}[1]{\mathbf{#1}}
\newcommand{\cpdist}[2]{p(#1|#2)} 
\newcommand{\info}{{\mathfrak I}}

\newcommand{\stateS}{\Omega({\mathcal H})}

\newcommand{\be}{\begin{enumerate}}
\newcommand{\ee}{\end{enumerate}}
\newcommand{\beq}{\begin{equation}}
\newcommand{\eeq}{\end{equation}}
\newcommand{\beqx}{\begin{displaymath}}
\newcommand{\eeqx}{\end{displaymath}}
\newcommand{\beqa}{\begin{eqnarray}}
\newcommand{\eeqa}{\end{eqnarray}}
\newcommand{\beqax}{\begin{eqnarray*}}
\newcommand{\eeqax}{\end{eqnarray*}}

\newcommand{\field}[1]{{\mathbb #1}} 
\newcommand{\cali}[1]{{\mathcal #1}}

\begin{document}
\title{Estimating Information Gain in Measurements in Suboptimal Bases for Quantum State Tomography}

\author{
Manas Patra \\
CQCT and Department of Computing, Macquarie University,\\ NSW 2109,
Australia\\ 
{manas@ics.mq.edu.au} 
\commentout{
\and 
Igor Shparlinski \\ 
Department of Computing, Macquarie University,\\ NSW 2109,
  Australia\\ 
{igor@ics.mq.edu.au}}
}

\date{} 
\maketitle 

\begin{abstract}
It is known that mutually unbiased bases, whenever they exist, are optimal in an information theoretic sense for the determination of 
unknown state of a quantum ensemble. These bases may not exist in most dimensions and some suboptimal choices have to be made. The present paper deals with
estimates of the information loss in suboptimal choice of bases. The information is calculated directly in terms of transition
probabilities. I give estimates for the information content of measurement in some approximate MUBs proposed recently. 

 \footnote{A shorter version of this paper appeared in the proceedings of EQIS 2005}
\commentout{
 }   
\end{abstract} 

\section{ Introduction} 
The state of a quantum system is completely specified by a ray in a
complex Hilbert space $\cali{H}$, or more generally by a density
matrix. A density matrix is a positive operator on $\cali{H}$ with
unit trace. Thus, a density matrix has nonnegative eigenvalues whose sum equals 1. In 
particular, it is hermitian. The Hilbert space $\cali{H}$ is in general
infinite dimensional. However, if we confine our attention to some
physical quantities like spin or polarisation then the corresponding
space is finite dimensional.  The complete space of the system 
is the tensor product of this finite dimensional space and an infinite
dimensional space which correspond to physical quantities like
momentum, energy, angular momentum etc. As long as the interaction
between these two types of quantities is negligible we may consider
them separately since the complete state is product or ``unentangled''
state. Henceforth, we will consider finite dimensional spaces
mostly. The finite dimensional case is of paramount interest in
quantum computing and information. 

Often the state of the quantum system is not known and has to be
determined by certain tests. For this, we need an ensemble. Imagine
for example, a preparation apparatus designed to prepare quantum
systems in some arbitrary dimension $n$ (qunits!) in some specified
pure state. Noise in the apparatus will distort the qunits and what we
will get is distribution over pure states: a density
matrix. Similarly, in the context of tomographic quantum
cryptography\cite{REK04}
we have to choose a set of positive-operator valued measures( POVM) s
for measurement. What is the optimal 
choice of such POVMs? This has been answered for projection valued
measures( PVM) by an information theoretic analysis\cite{FW89}. Let us distinguish 
two problems concerned with general measurements. The first is the problem of estimation 
or hypotheses testing \cite{Helstrom}: given a measurement procedure and the data 
find the best estimates of the state that produced the data , assuming some prior 
distribution on the state. A recent thorough analysis may be found in \cite{Hayashi}. The 
second problem may termed as a {\em design} problem. Given a class of measurement to determine 
the parameters characterizing the state, find the "best measurement". Now, the best measurement will most likely depend upon the state, so we look for the best measurement on 
the average. That is we optimize the average information gain corresponding to the different 
measurements from the given family. The present paper is concerned with the second problem.

Let $\cali{H}$ be a Hilbert space of dimension $n$. Let $V(\cali{H})$
be the set of hermitian operators on $\cali{H}$. The dimension of
$V(\cali{H})$ as a {\em real} vector space is $n^2$. Let
$\Omega(\cali{H})\subset V(\cali{H})$ be the set of positive operators
with trace 1. It is a convex set. The map \( L:V(\cali{H}) \rightarrow
V(\cali{H}) \text{ such that } L(T) =T - 1/n \tr(T)\cdot I\) is linear
and the image $l(\cali{H})$ is the space of hermitian operators with
trace 0. It has dimension $n^2-1$. Here $I$ denotes the identity
operator. The affine space $l(\cali{H})+I/n$  consists of all hermitian
operators with trace 1. Therefore, a density matrix is completely
specified by $n^2-1$ parameters. 

The state of a quantum system is not directly measurable. A
measurement yields only probabilities. We assume for
simplicity that all hermitian operators are observable. Thus, let 
$\rho\in  \stateS$ is a state and $A$ be hermitian operator. Let $A=
\sum_i c_i\pj{\alpha_i}$ be the spectral decomposition of $A$. Here we
use the familiar Dirac notation: $\bra{\alpha}$ is the unique dual
vector (via the inner product) of the vector $\ket{\alpha}$ in
$\cali{H}$. A measurement of $A$ will record one of the eigenvalues
$c_i$ with probability $\tr(\pj{\alpha_i}\rho) =
\inpr{\alpha_i}{\rho}{\alpha_i}$  The space of linear operators on
$\cali{H}$ becomes a Hilbert space of complex dimension $n$ by
defining the inner product 
\beq \label{eq:FrobNorm}
\langle B,C\rangle = \tr(\hconj{B}C). 
\eeq    
The corresponding norm is the Frobenius norm. Its restriction to
$V(\cali{H})$ makes the latter a real inner product space of
dimension $n^2$. Let us assume that $A$ is nondegenerate. The
probability of obtaining the $i^{th}$ outcome is 
\beq \label{eq:probProj}
p_i= \tr(\pj{\alpha_i}\rho) = \inpop{\pj{\alpha_i}}{\rho}.   
\eeq      
The probabilities may be interpreted as projections of the state 
$\rho$ onto the corresponding ``coordinate'' vector
$\pj{\alpha_i}$. Of course, here a vector means an element in
the space $V(\cali{H})$. It is more convenient to consider the
traceless hermitian operators $\rho -I/n$ instead of $\rho$.
In any case a measurement on an ensemble in some basis can at best
give us an 
estimate of the $n^2$ probabilities of the possible outcomes. We may
thus characterize a measurement of a nondegenerate observable $A$ by
the corresponding orthonormal basis in the spectral decomposition. We
shall henceforth simply refer to the bases of measurement. Of the
$n^2$ probabilities obtained by measurement in some basis $\cali{B}$
only $n^2-1$ are independent since $\sum_i p_i=1$. Let $P_i=
\pj{\alpha_}$ projection operators corresponding to $\cali{B}$. They
satisfy $P_iP_j = \delta_{ij} P_i$ and $\sum_i P_i= I$. As vectors in
$V(\cali{H})$ they are linearly independent. To avoid confusion we
call the projection operators corresponding to some basis in the
ambient space $\cali{H}$ projection {\em vectors} or simply {\em
  projectors} when they are considered as elements of $V(\cali{H})$.  
But there are only $n$ of them. Further, if we have two bases
$\cali{B}_i \text{ and } \cali{B}_2$  then at most $n-1$ projectors
from $\cali{B}_2$ can be independent of those in $\cali{B}_1$ due to
the relation $\sum_i P_i= I$. Hence to get the $n^2-1$ coordinates of
$\rho$ we need $n+1$ bases such that they are independent in the
following sense. Pick the first $n-1$
projectors each from each of the basis. If they are independent 
 after the affine transformation described
above  they constitute a basis in $l(\cali{H})$. We call such a set of
projectors a complete set of measurement bases( CSMB for
short). Suppose we have two CSMBs $\cali{S}_1 \text{ and }
\cali{S}_2$. If all other conditions remain same which one should we
pick for determining the unknown state of a quantum ensemble? We may
assume ideal conditions- perfect preparation 
procedures, perfect detectors and measuring devices etc.- to compare
the two. In a classic paper \cite{FW89} Fields and Wootters proved
that a set of mutually unbiased bases(MUBs) is an optimal choice. Two
sets of orthonormal bases \( \{\ket{\alpha_i}\}\text{ and }
  \{\ket{\beta_j}\} \)are said to be mutually unbiased if
  $|\inpr{\alpha_i}{\beta_j}|^2= 1/n$.    
They further go on to show that such bases exist whenever the
dimension $n$ 
is a prime power, extending earlier work of Ivanovic\cite{I81} who
proved the existence of MUBs in prime dimension. These works however
left open the question of the existence of MUBs for $n$ which divides
two or more distinct primes e.g.\ 6. It is now widely believed that
MUBs do not exist in such dimensions. However, we can expect CSMBs
which approximate MUBs. Then it is natural to ask: how much do we lose
in the approximation process. This question is  relevant even in the
cases where MUBs are known to exist because in more realistic
situations the measurement apparatus will only approximately implement
the MUBs. However, to answer such questions we must have an
appropriate framework in which these questions may be posed and
answered precisely and quantitatively. The natural candidate seems to
be information theory. 

In this work I elaborate on the information content of a quantum
measurement process. This was partly done in \cite{FW89}. I then give
estimates for the information 
content of measurements in CSMBs which approximate MUBs. Even in the
case where MUBs are known to exist there is always a margin of
error. So even here it is reasonable to estimate the information
content of CSMBs. 

I point out that information optimization appears in a different context in estimation theory, namely, hypothesis testing. Given some prior information about the distribution of states and the outcomes of some experiment we seek for optimal choice of state form the experimental data. `

\section{Information content of a measurement}  
In this section we follow \cite{Lind56} to define the information
content of a measurement and apply it to the case of measurements for
the determination of the quantum state of an ensemble. Let $\cali{M}$
be a measurement on some system $S$. We should use the term
``experiment'' rather than measurement since the latter seems to imply
a single measurement. Let $S$ be characterized by some parameters
denoted by $\theta$ which will usually be drawn from some subset
$\rvec{\Theta}$ of
$\field{R}^k$, the $k$-dimensional Euclidean space. Let $p(\theta)$
represent the {\em a priori} probability distribution of the
parameters $\theta$. Corresponding to every value of $\theta$ there is
a probability measure on $\rvec{X}$- the set of possible measurement
data which is again a subset of some Euclidean space. We assume
that this measure is given by $\cpdist{\rvec{x}}{\theta}d\rvec{x}$. 
$\int_B\cpdist{\rvec{x}} {\theta}d\rvec{x}$ is the  conditional 
probability of getting the outcome $\rvec{x}$ in $B\subset \rvec{X}$
given the state  $\theta$. Let $\pdist{\rvec{X}}=\int_{\rvec{X}}
\cpdist{\rvec{x}} {\theta} \pdist{\theta} d\theta$ be the probability
density of the random variable $\rvec{x}$. Note that, we have used the
same symbol $p$ for the probability densities of different random
variables. This does not of course imply that they are the same
functions. The notation is more convenient and unambiguous if taken in
proper context. Moreover, we do nit differentiate between a random
variable and its values.  In an experiment we often are often
interested in the {\em posterior} probability 
$\cpdist{\theta}{\rvec{x}}$. That is, given the measured values
$\rvec{x}$ the probability density for $\theta$ which in turn gives us
the probability of the state. This is the primary problem in
estimation theory and hypothesis testing. The information content of
the measurement $\cali{M}$ is defined as 
\beq \label{eq:defInfo}
\info(\cali{M}, p(\theta) ,\rvec{x})\equiv \int
\cpdist{\theta}{\rvec{x}} \log \cpdist{\theta}{\rvec{x}}\, d\theta -
\int p(\theta)\log p(\theta)\, d\theta.
\eeq 
If $\cpdist{\theta}{\rvec{x}}=0$ then the integrand is defined to be
zero and the logarithm is taken over an arbitrary but fixed base. The
justification for this definition is as follows. Consider the term 
\[ \info_0 \equiv \int p(\theta)\log p(\theta)\, d\theta. \]
It is supposed to represent the prior information about the state
$\theta$. Let us take a simple example to illustrate an important
property. Suppose it is known that the state $\theta$ is found in  
$\rvec{\Theta}'\subset \rvec{\Theta}$ with probability $q$. Let
${\mathcal I}_1$ be a measure of information corresponding to the
knowledge whether $\theta$ is in $\rvec{\Theta}'$ or its
complement. Let ${\mathcal I}_2$ and ${\mathcal I}_3$ be the amount of
information gained in the next phase when get the value of $\theta$ in
$\rvec{\Theta}'$ or its complement respectively. Then a fundamental
additive property required of the information measure is that the
total information 
\beq
{\mathcal I} = {\mathcal I}_1 + q{\mathcal I}_2+  (1-q){\mathcal I}_3
\eeq
Then it is not difficult to show that the information measure
$\info_0$ is unique up to a constant multiple. We do not discuss these
points further but refer the reader to any good source on basic
information theory e.g.\ \cite{Khin} and \cite{Lind56} for a
discussion in the context of experiments. The difference between the
posterior information $\info_1 = \int\cpdist{\theta} {\rvec{x}} \log
\cpdist{\theta}{\rvec{x}}\, d\theta$ and the prior information
$\info_0 =  \intp(\theta)\log p(\theta)\, d\theta $ is the net
information gain. It depends upon the experiment and the distribution
of the data $\rvec{x}$. Thus we may say that one
experiment or measurement is more informative than other. Let us
calculate information content for some simple measurements in the
quantum domain. Let the dimension $n=2$. Suppose we have prior
information that the state is a pure state $\ket{0}$ or $\ket{1}$ with
probability 1/2. We may therefore model the parameter space as
$\rvec{\Theta} = 
\{0, 1\}$ with $p(0)=p(1)=1/2$. Then $\info_0 = 1/2\log {(1/2)} + 
1/2\log {(1/2)}= -1.$ The logarithm is taken to the base 2. 
Now suppose that we choose to make measurement $\cali{M}_1$ in
the basis $\{\ket{0}, \ket{1}\}$ which is natural, given the prior
information. Then the conditional probabilities may be conveniently
written in the matrix form, for $i,j\in \{0,1\}$

\[p(i|j) = 
\begin{pmatrix} 1 & 0 \\ 0 & 1 \end{pmatrix} \]

It is simply the unit matrix of order 2. Thus, if we get the
measurement outcome 0 we are sure that the state of the system was
$\ket{0}$ etc. Then it is easy to see that $\inf_1 (\cali{M}_1, i) =
0$ and hence $\info (\cali{M}_1, i)= \info_1(\cali{M}_1, i)- \info_0=
1$. Now suppose we perversely 
choose the basis $\ket{\stackrel{+}{-}}=
\frac{1}{\sqrt{2}} (\ket{0}{\stackrel{+}{-}} \ket{-})$ for measurement
$\cali{M}_2 $. Then the corresponding conditional probability matrix is 
 \beqx
p(i|j) = \begin{pmatrix} 1/2 & 1/2 \\ 1/2 & 1/2 \end{pmatrix} 
\eeqx 
Again it is easy to see that the information gain in this case is
$\info (\cali{M}_2, i)=0$. That is we get no information from
$\cali{M}_2$. In fact, the average information, to be defined below,
is  zero. This is of course intuitively obvious from the choice of
basis in $\cali{M}_2$. 

The information measure defined above depends on the 
state and may be negative. But the average information 
\beq 
\info(\cali{M}, p(\theta)) \equiv \int \info(\cali{M}, p(\theta)
,\rvec{x})p(\rvec{x}) d\rvec{x}  
\eeq  
is independent of the state and is nonnegative \cite{Lind56}. Here,the
probability density 
\beq
p(\rvec{x}) = \int p(\rvec{x}|\theta) d\theta
\eeq 
is the mean probability distributions averaged over $\theta$. 
It is not difficult to show that 
\beq \label{eq:avInfo}
\info(\cali{M}, p(\theta)) = \int \int p(\theta)p(\rvec{x}|\theta) \log
     {(p(\rvec{x}|\theta ))}d\rvec{x}d\theta -\int p(\rvec{x})
     \log{(p(\rvec{x}))}d \rvec{x} 
\eeq
This is the formula we use estimate the information gained in quantum
measurements. 
\section{Quantum state tomography and MUBs} 
Given an n-dimensional quantum ensemble in an unknown state how do we
determine its state? This is the problem of quantum state
tomography. The state is not directly observable but we may infer it 
from the probability distributions observed. As mentioned in the
Introduction we need (projective) measurement in $n+1$ bases to
determine the state completely from the observed
probabilities. Actually, the state tomography problem has broadly two
theoretical aspects. The first is a design issue. What is the optimal
choice of bases? The second aspect is a problem of decision or
estimation theory: for a given measurement what is the best possible
estimate of the parameters characterizing the state? In this paper we
will be mainly concerned with first aspect. So let us formulate the
problem precisely now. 

Given an ensemble of quantum systems in some unknown state $\rho$. By  
an ensemble we mean an unlimited supply of identically prepared
quantum systems. Let $\cali{B}^1, \ldots , \cali{B}^{n+1}$ be $n+1$
base in $\cali{H}$ with 
\beq \label{def:basisPVM}
\cali{B}^k = \{\field{P}_i^k \equiv \pj{ \alpha_i^k}\}_{i=1}^{n} 
\eeq  
As vectors in the $n^2$-dimensional space $V(\cali{H})$ at most $n^2$
of them can be linearly independent. Due to the relations $\sum_i
\field{P}_i^k = I $ for all $k$ we can only have all the $n$ vectors
from exactly one basis in any independent set and if $\cali{B}=
\bigcup_k \cali{B}^k$ contains a maximal independent set then we
choose the first $n-1$ projectors $\field{P}_i^k, \, i= 1\cdots n-1$
and $\frac{I}{n}$ as a basis for $V(\cali{H})$. Let $s^{kl}_{ij} =
\langle \field{P}_i^k,  \field{P}_j^l\rangle = \tr( \field{P}_i^k
\field{P}_j^l)$. The nonnegative numbers $s^{kl}_{ij}$ are the
respective transition probabilities among the vectors in the $k^{th}$
and $l^{th}$ basis. Note that $s_{ij}^{kk} = \delta_{ij}$ since each
of the basis is orthonormal. It was seen in Section 1 that
$\{\field{P}_i^k -I/n: \; i=1,\ldots, n-1 \text{ and } k=1,\ldots, n+1\
$ form a basis for $l(\cali{H})$, the space of traceless hermitian
operators. 
\commentout{
But no projector can be orthogonal to
{\em all} the projectors in another basis. Let 
\beq \label{eq:stateParam}
\rho = aI/n + \sum_{i,k}^{\substack{k=n+1 \\ i=n-1}}y_i^k
\field{P}_i^k 
\eeq
Since $\tr(\rho)=1$ this equation may be written as 
\beq
\rho -I/n= \sum y_i^k (\field{P}_i^k -I/n)\equiv  \sum y_i^k T_i^k  
\eeq 
}
Thus, for a state $\rho$ let 
\beq
\rho -I/n= \sum y_i^k (\field{P}_i^k -I/n)\equiv  \sum y_i^k T_i^k  
\eeq 
Then, 
\beq \label{eq:basicProb} 
\tr((\rho-I/n) T_j^l) = p_j^l -1/n= \sum y_i^k \langle T_i^k ,T_j^l
\rangle = \sum_{i,k} t_{ij}^{kl} y_i^k 
\eeq 
It is easy to see that if the original bases are mutually unbiased
then $\langle T_i^k, T_j^l\rangle = 0\text{ for } k\neq l$, that is
the operators $T_i^k  \text{ and } T_j^l, \; k \neq l$ are orthogonal
when considered as vectors.  
Here 
\beq \label{eq:defT}
t_{ij}^{kl} \equiv \langle T_i^k ,T_j^l \rangle = \tr(T_i^k
T_j^l)= s_{ij}^{kl} -1/n
\eeq
and $p_i^k$ is the probability of $i^{th}$
outcome in the measurement in the $k^{th}$ basis. If we consider the
parallelepiped spanned by the vectors $T_i^k$ then $t_{ij}^{kl}$ are the
angles between the sides  $T_i^k \text{ and } T_j^l$. Notice also that
the input parameters characterising the state (denoted by $\theta$
earlier) are the components $y_i^k$. We will denote these by a vector
$\rvec{Y}$. 

By a measurement we mean a collection of several observations
in different bases on subensembles of the original ensemble. We
picture a massively parallel 
setup where we have a several measuring devices $\cali{D}_k$  for each
basis $\cali{B}^k$. Th original ensemble is divided into large
subensembles and tested by each of these $n+1$ groups of devices. For
each $k\leq n+1$ we get frequencies $m_i^k$ for the $i^{th}$ outcome,
$1\leq i \leq n-1$ in the  $k^{th}$ device group. The numbers $m_i^k$
constitute the measurement data $\rvec{x}$ in  \ref{eq:defInfo} and
\ref{eq:avInfo}. What is a reasonable probability distribution for the
$m_i^k$? Here we appeal to the local limit theorem \cite{Gneden} in
probability theory which roughly states that for independent discrete
random variables the probability distribution of their frequencies
tends to the normal distribution in the limit $N\rightarrow \infty$,
$N$ the number of trials.  
\commentout{
We do not state the precise version of the theorem
but make the following theorem in the context of our problem which
holds asymptotically. 
\begin{thm}
With the notation as above let $P(m_i^k|\rvec{Y}),\, i=1, \ldots, n-1
\text{ and } k= 1, \ldots , n+1$ denote the probability of occurrence
of the corresponding outcome $m_i^k$ times. Then

\beq \label{eq:basicAssum} 
P(m_i^k|\rvec{Y}) \rightarrow
\frac{\exp{(-1/2\sum_{i,k}\frac{(m_i^k - Np_i^k )^2} { Np_i^k})}}
     {(2\pi N)^{\frac{n^2-1} {2} } \sqrt{\prod_{ik}p_i^k}}   
\eeq 
\end{thm} 
where $N$ is the number of trials. 
We will assume the function on the right side as the
distribution function for the $m_i^k$, conditioned by the parameters
$\rvec{Y}$ characterising the state. There is a technical point. From
the expression we see that this is defined only when none of the
$p_i^k$ are 0 or 1. The set of states $\rho$ which give probability 0
or 1 has measure zero. Hence, the equation is defined almost
everywhere. Since we are going to integrate it over the whole space we
may define $P(m_i^k|\rvec{Y})$ arbitrarily at these points. The
limiting distribution is over all the $n$ outcomes that are possible
in a measurement. The frequencies must satisfy 
\[ \sum_{i=1}^n m_i^k = N \]
}
We must have some prior distribution for the states. Let $V$ be the
volume of the parallelepiped spanned by the vectors $T^k_i$. 
Assuming a
uniform distribution for the states it can be shown that\cite{FW89}
that the information gain in a quantum test is proportional to
$\ln{V}$ apart from an additive constant. We will in fact take
$\ln{V}$ as the measure for information content of a quantum test of
an ensemble in CSMB and for a CSMB $\cali {C}$ write
$\cali{I}(\cali{C})$  for the information gain and $V(\cali{C})$ for
the corresponding volume. The first result I prove was already given in
\cite{FW89} but the present approach is different. 

\begin{thm} \label{thm:maxInfo}
Information gain $\cali{I}(\cali{C})$ is maximum if and only if
$\cali{C}$ consists of mutually unbiased bases. 
\end{thm}

\noindent
{\bf Proof}. From the preceding discussion, we have to show that the
  volume $V(\cali{C})$ spanned by the vectors $T_i^k$ is maximal iff
  $T_i^k$ and $T_j^l$ are orthogonal for $k\neq l$. Notice first that
  $\inpop{T_i^k}{T_j^k} = \delta_{ij} - 1/n$. Consider the
  $(n^2-1)\times (n^2-1)$ matrix $\Gamma(\cali{C}) = (t_{ij}^{kl}) =
  \inpop{T_i^k}{T_j^k}$ and assume the ordering defined by the pair
  $\{k,i\}$. This simply means that the matrix consists of $n+1$
  blocks $\gamma^{kl}$, each a square matrix of size $(n-1)$   such
  that $ \gamma^{kl}(ij) =t_{ij}^{kl}$. 
\beq
\Gamma(\cali{C}) = \begin{pmatrix} \gamma^{11} & \gamma^{12}
  & \ldots & \gamma^{1n+1} \\ 
\gamma^{21} & \gamma^{22} & \ldots & \gamma^{2n+1} \\ 
\vdots & \vdots  & \ldots & \vdots \\ 
& & \ldots & \gamma^{n+1,n+1} 
\end{pmatrix} 
\eeq 
If we choose any orthonormal basis for $l(\cali{H})$ 
and express $T_i^k$ in this basis. Let $\cali{T}$ be the corresponding 
real matrix of the coefficients then it is clear that
$\cali{T}\cali{T}^t = \Gamma (\cali{C})$, where $A^t$ is the transpose
of $A$. It follows that $\det{\Gamma (\cali{C})}= (V(\cali{C}))^2$ and
$\Gamma (\cali{C})$ 
is positive definite. Thus maximizing 
$V(\cali{C})$ is equivalent to maximising $\Gamma (\cali{C})$. Below we
will focus on the latter. From the generalised Fischer-Hadamard
inequality \cite{ES76} it follows that 
\beq \label{ineq:Fischer} 
\det{\Gamma(\cali{C})} \leq \det{\gamma^{11}}\cdots \det{\gamma^{n+1,n+1}}
\eeq
The rhs is determinant of the product of the diagonal blocks in
$\Gamma{\cali{C}}$. Now, if the $T_i^k$ are orthogonal then the
off-diagonal blocks are all zero matrices and the  equality holds in
eq.(\ref{ineq:Fischer}). This proves the sufficiency part. 

The equality holds in \ref{ineq:Fischer} only if the following
condition is satisfied \cite{ES76}. Let $S$ be the $(n+1)\times (n+1)$
matrix such that $S(ij) = 1$ if $\gamma^{ij}\neq 0$ and $S(ij)= 0$
otherwise. Then the equality holds if and only there is permutation
matrix of order $n+1$ such that $PSP^{-1}$ is triangular. Since
$\Gamma$ is symmetric and $P^{-1}=P^t$ it follows that if $PSP^{-1}$
is triangular it must be diagonal. The operation $S \rightarrow
PSP^{-1}$ permutes the diagonal elements of $S$ among
themselves. Hence, $PSP^{-1}$ is diagonal iff all off-diagonal
elements are zero. That is, $\gamma^{ij} =0 \text{ for } i\neq
j$. That is the original bases are mutually unbiased. The necessity is
proved.    

The above theorem gives an upper bound. A natural question is: how
tight is the bound. This is related to the estimation of the
information content in bases which are complete but not mutually
unbiased. We now give an estimate of the relative loss due to such a
non-optimal choice. First let us compute the determinant in the case
of MUBs. We only have diagonal terms. Recall that a diagonal block
$\gamma^{kk}(ij) = \inpop{\field{P}_i^k- 1/n}{\field{P}_j^k-1/n}=
1-1/n$. We write this as $\gamma^{kk} = I-1/nT$, $I$ is the identity
matrix of order $n-1$ and $T$ is the matrix with all entries 1. Let
$\Gamma_0$ be the submatrix of $\Gamma$ consisting of the diagonal
blocks. Note that all the diagonal blocks $\gamma^{kk}$ are
identical. 

\begin{lem}
\(\det{\Gamma_0} = \frac{1}{n^{n+1}}\). 
\end{lem}

\noindent
Proof. First note that $T^2=(n-1)T$. The eigenvalues of $T$ are
therefore, $n-1$ and 0. The rank of $T$ is 1. Hence the eigenvalues of
$I-1/nT$ are $1/n$ and 1. The determinant of each block is therefore
$1/n$ and since there are $n+1$ blocks the result follows. 

\noindent 
If $\cali{C}$ is MUB then $\Gamma(\cali{C})$ has all off diagonals
zero. Now $\gamma^{-1}= (I+1/nT)^{-1}= I+T$. Hence in block form we
have, 
\beq \label{eq:compDet1}
\det{\Gamma(\cali{C})}= \det{\Gamma_0} \cdot \det{\begin{pmatrix} I &
    (I+T)\gamma^{11} & \cdots \\ \vdots & \vdots & \vdots \\ \cdots &
    \cdots & I   \end{pmatrix}}
\eeq
 That is, the off-diagonal blocks are multiplied by the matrix
 $I+T$. Consider $\gamma^{kl}$. Recall that 
 $\gamma^{kl}(ij) = s^{kl}_{ij}-1/n, \; i,j \leq n-1$, where
 $s^{kl}_{ij}$ are transition probabilities. An easy calculation shows
 that $(I+T)\gamma^{kl}(ij) = s^{kl}_{ij} - s^{kl}_{in}$. The
 term $s^{kl}_{in}$ appear because we omitted the $n^{th}$
 basis vector from each basis in the state space $\cali{H}$. If we had
 chosen another vector, say, the first then $s^{kl}_{i1}$ would have
 been subtracted. The point is the information content depends on the
 {\em differences} of probabilities. Only in the case of MUBs are these
 differences all zero. Next we give an estimate in the general case. 
\begin{thm}\label{thm:estDet1} 
Let $|s^{kl}_{ij} - s^{kl}_{ir}| < \varepsilon $ for some $\varepsilon>0$. Let
$\Gamma' =   \Gamma(\cali{C})- I_{n^2-1}$ and let
$\lambda_m$ be the 
  minimum eigenvalue of $\Gamma'$. Then  
\beq \label{eq:estDet1}
   e^{\frac{(n^2-n)^2(n^2-1)\varepsilon^2}{1+\lambda_m}}
   \frac{\det{\Gamma(\cali{C}})} {\det{\Gamma_0}} \geq 1  
\eeq
\end{thm}

\noindent
Proof. The theorem is a direct consequence of an estimate given in
\cite{Ipsen}. From its definition $\Gamma(\cali{C})$ is positive
semidefinite because it is a real matrix of the form $<{\tt
    b}_i, {\tt b}_j>$ for vectors ${\tt b}_i$ in appropriate
dimension. Hence the estimate in \cite{Ipsen} 
is applicable. The upper bound is just the Hadamard-Fischer
inequality. The lower estimate in \cite{Ipsen} is
$e^{\frac{-(n^2-1)\rho^2} {1+\lambda_m}}$, where $\rho= max\{|\lambda_i|
:\; \lambda_i \text{ an eigenvalue } \}$ is the spectral radius of
$\Gamma'$. The fact that $\rho \leq max \{ |R_i|\}$, where $|R_i|$ is
the sum of absolute values of the entries in ith.\ row of
$\Gamma'$ is easily proved\cite{MM}. We get $n^2-n$ because the
diagonal blocks in $\Gamma'$ are zero.  

Let $v_d \equiv \frac{\det{\Gamma{\cali{C}}}} {\det{\Gamma_0}}$. 
As an illustration let $\varepsilon \leq 1/n^4$ then a simple
calculation yields $ \det{\Gamma(\cali{C})}/\det{\Gamma_0} \geq
e^{-1/n^2} $ and the corresponding loss in information is 
$O(1/n^2)$. In the cases where MUBs are known to exist, that is when
$n$ is a prime power it is natural to expect that in some actual
designing for testing in these bases there would be errors. If we can
bound give an estimate $\varepsilon$ for these errors then the
information loss can be estimated. Even in cases where MUBs are not
known to exist approximate MUBs may be constructed
\cite{KRSW}. However, a direct application of the above estimates to
their constructions does not yield very good lower bounds. If
$\varepsilon \leq 1/n^3$, as in some cases of \cite{KRSW}, then  
the information loss can be estimated to be less than
$a=o(1)$. We now give an exact calculation of the determinant in the 
 second construction in \cite{KRSW}. 
 \commentout{That is, the
corresponding projection 
operators are linearly {\em dependent}. However, one may do
``unitary'' perturbation to ensure independence.}
\subsection{Calculation of determinants in special cases} 
In the case of KRSW
construction 
\beq \label{eq:transiProb}
\begin{split}
 s^{ab}_{ij} = & \delta_{ij},  \quad a= b \\ 
             = & \frac{n+1}{n^2},  \quad a\neq b, i\neq j \\
             = & \frac{1}{n^2},  \quad a\neq b, i=j 
\end{split} 
\eeq
We calculate the determinant of the $(n^2-1) \times (n^2-1)$ matrix
defined by the numbers \(t^{ab}_{ij}=s^{ab}_{ij}, \; a,b=1, \ldots,
n+1 \text{ and 
} i,j = 1, \dots, n-1 \). The rows(columns) of the matrix are indexed
by pairs $[a,i]([b,j])$. We do it for a slightly more general case.

It is clear that $s^{aa}_{ij}= \delta_{ij}$. Let $\Gamma$ be the
matrix whose entries in block form are the $(n-1)\times (n-1)$
matrices $\gamma{ab}$ where 
\[ \gamma^{ab}_{ij} = t^{ab}_{ij} \] 
Let $D$ be the matrix which contains only diagonal blocks. Thus,
\beqx
D= \begin{pmatrix} \gamma^{11} & 0 & \cdots & 0 \\
0 & \gamma^{22} & \cdots & 0 \\
0 & \vdots & \cdots & 0 \\
0 & \cdots & 0 & \gamma^{n-1,n-1} 
\end{pmatrix} 
\eeqx 
Then one can show that $\det{\Gamma} = \det{D}\cdot \det{\Gamma'}$
where 
\begin{gather} 
\Gamma' \equiv  \begin{pmatrix} I & \Psi^{12} & \cdots & \Psi^{1,n+1}\\
0 & I & \cdots & 0 \\
0 & \vdots & \cdots & 0 \\
0 & \cdots & 0 & \gamma^{n-1,n-1} 
\end{pmatrix} 
\quad \Psi^{ab}_{ij} = s^{ab}_{ij} - s^{ab}_{nj}, \; i,j = 1,\ldots,
n-1 
\end{gather} 
and $I$ is the unit matrix of order $n-1$. The apparent lack of
symmetry in $\Gamma'$ can be removed by successively subtracting the
$(i+1)^{th}$ row from the $i^{th}$ in each row of blocks. The result
is that typical entries are of the form 
\[ s^{ab}_{ij} - s^{ab}_{i+1,j} \] 
However, for the present purpose we stay with the first form of
$\Gamma'$. We calculate the determinant for a very special case. Let 
\beq \label{eq:2-value} 
s^{ab}_{ij} = \begin{cases} 
c+\frac{1}{n}& \text{if $a\neq b$ and $i\neq j$ }\\
\frac{1}{n} -(n-1)c & \text{if $a\neq b$ and $i=j$ }\\
\delta_{ij} & \text{if $a=b$}. 
\end{cases}
\eeq 
Then each of the off-diagonal blocks in $\Gamma'$ is diagonal, 
\begin{gather}
\Psi^{ab} \equiv A = \begin{pmatrix} -nc & 0 & \dotsb & 0 \\
              0 & \ddots & \dotsb & 0 \\
              0 & & \dotsb & -nc 
\end{pmatrix}  \text{ and } \\
 \Gamma' = \begin{pmatrix} I &  A & \dotsb & A \\
               A & I & \dotsb & A \\ 
               \vdots & \vdots & & \vdots \\
              A & A & \dotsb & I  
\end{pmatrix}
\end{gather}
First, we calculate the eigenvalues of $\Gamma'$ together with their
multiplicities. This will give us the determinant. Let 
\[ X = \begin{pmatrix} {\tt x}_1 \\ \vdots \\  {\tt x}_{n+1} \end{pmatrix} \] 
where \(X\) is a $(n^2-1)$ column vector and ${\tt x}_i$ are
$(n-1)$ column vectors. This decomposition is made to
match the decomposition of the matrix $\Gamma'$. Then if $X$ is an
eigenvector with eigenvalue $d$, 
\[\Gamma'X = dX \imp {\tt x}_i -nc \sum_{j\neq i}^{n+1} = d{\tt
  x}_i,\; i= 1, \dotsc , n+1 \]
Let ${\tt y} = \sum_{i=1}^{n+1} {\tt x}_i$ be a $n-1$ column
vector. Then the above equation can be written as 
\[ (1+ nc-d) {\tt x}_i = nc {\tt y} \] 
Consider two possibilities. First, ${\tt y}=0$. Then, $d=
1+nc$ pro. Now, the subspace of $\field{C}^{n^2-1}$
corresponding to the solutions ${\tt y}=0$ is $n^2-n$
dimensional, equal to the multiplicity of the eigenvalue $d=
\frac{1}{1+nc}$. The second case is when ${\tt y}\neq 0$. Then clearly
all the $(n-1)$-vectors are equal, i.e.\ , 
\[ {\tt x}_i = \frac{nc}{1+ nc-d} {\tt y} =  \frac{n(n+1)c}{1+ nc-d}
    {\tt x}_i \]
Hence, in this case $d= 1-n^2c$. The multiplicity is clearly
			     $n-1$. Thus, 
\[ \det{\Gamma'} =\left(1+nc \right)^{n^2-n}
\left(1-n^2c \right)^{n-1} \] 

For example, if $c= 1/n^2$ then the determinant is zero. This is the
case in [KRSW05]. However, notice that the main contribution to the
determinant comes from the eigenvalue $1+nc$ but remembering that \(
\det{\Gamma'} \leq 1 \). This places restriction on $c$.     

\section{Discussion} 
We analysed the information content of a quantum state tomography in
PVMs. The information seems to depend upon the choice of the basis
vector we eliminate initially in constructing $T_i^k$ (see discussion
following eq.(\ref{def:basisPVM})), in this case, $\cali{B}^i_n
$. However, it is easy to see that the information measures
corresponding to different 
  choices differ by an unimportant additive constant. Another point is
  that the parallelepiped whose volume was used as a measure for
  information is slightly different from the one given in
  \cite{FW89}. But again the corresponding information measures differ
  by an additive constant. What is perhaps more important
  and difficult is the weakening of assumption of uniform prior
  distribution and a characterization of the optimal 
  PVM as a {\em function} of the prior distribution
  function. Another related issue is a similar investigation of
  optimal POVMs. Two other difficult problems are:1.\ estimating  
  information content of measurements on infinite dimensional systems
  and 2.\ incomplete measurements. These issues will be discussed in a
  forthcoming paper.  

\bibliographystyle{phaip}

\begin{thebibliography}{10}

\bibitem{REK04}
J.~\={R}eh\'{a}\={c}ek, B.-G. Englert, and D.~Kaszlikowski,
\newblock Phys. Rev. A {\bf 70}, 052321 (2004).

\bibitem{FW89}
W.~K. Wootters and B.~D. Fields,
\newblock Annals of Physics {\bf 191}, 363 (1989).

\bibitem{Helstrom}
C.~W. Helstrom,
\newblock {\em Quantum estimation and detection theory},
\newblock Academic, New York, 1976.

\bibitem{Hayashi}
M.~Hayashi,
\newblock J. Phys. A {\bf 31}, 4633 (1998).

\bibitem{I81}
I.~D. Ivanovic,
\newblock J. Phys. A {\bf 14}, 3241 (1981).

\bibitem{Lind56}
D.~V. Lindley,
\newblock Ann. Math. Statist. {\bf 27}, 986 (1956).

\bibitem{Khin}
A.~I. Khinchin,
\newblock {\em Mathematical foundations of information theory},
\newblock Dover, 1957.

\bibitem{Gneden}
B.~V. Gnedenko,
\newblock {\em The theory of probability},
\newblock Chelsea, 1967,
\newblock pp 95-117.

\bibitem{ES76}
G.~Engel and H.~Schneider,
\newblock Lin. Multi. Alg. {\bf 4}, 155 (1976).

\bibitem{Ipsen}
I.~C.~F. Ipsen and D.~J. Lee,
\newblock Determinant approximation,
\newblock http://www.ncsu.edu/\~ipsen/, 2003.

\bibitem{MM}
M.~Marcus and H.~Minc,
\newblock {\em A survey of matrix theory and matrix inequalities},
\newblock Dover, 1992.

\bibitem{KRSW}
A.~Klappenecker, M.~Rotteler, I.~E. Shparlinski, and A.~Winterhof,
\newblock On approximately symmetric informationally complete povms and related
  systems of quantum states,
\newblock http://xxx.lanl.gov/abs/quant-ph/0503239, 2005.

\end{thebibliography}

\end{document}